# Determining Multifunctional Genes and Diseases in Human Using Gene Ontology

Hisham Al-Mubaid[1], Sasikanth Potu[1], and M. Shenify[2]
[1]Dept. of Computer Science. University of Houston - Clear Lake, Houston, TX 77058, USA
[2]University of Baha, Baha, KSA
[1]Email: Hisham@uhcl.edu

**Abstract**

The study of human genes and diseases is very rewarding and can lead to improvements in healthcare, disease diagnostics and drug discovery. In this paper, we further our previous study on gene–disease relationship specifically with the multifunctional genes. We investigate the multifunctional gene–disease relationship based on the published molecular function annotations of genes from the Gene Ontology which is the most comprehensive source on gene functions. We present a computational approach based on path length between molecular function annotations as our main metric for estimating the semantics of gene functions and multifunctionality in connection with gene–disease association. We utilized functional genomics data from OMIM, the Gene Ontology, and GOA Annotation databases and show that multifunctional genes based on *mf* path length approach are significantly more likely to be involved in human diseases. We conducted several evaluations and the results show that the path length (*shortest distance*) between gene *mf* annotations highly correlates with gene multifunctionality and gene – disease association. Our main contribution is the way we identify and determine multifunctional genes using GO *mf* path length which has not been used or investigated before as gene multifunctionality metric.

## 1. Introduction

The study of multifunctional genes and diseases in the human genome is a very rewarding work and can benefit other branches of science on the long term [1–5]. Multifunctional genes are those human genes that are associated with more than one function within the cell [1, 2]. It has been shown by numerous studies that such genes are more likely to be involved in diseases [1, 4, 8, 9]. Therefore, studying these multifunctional genes and furthering the knowledge about gene multifunctionality will help medicine and drug discovery. This paper is intended to further the knowledge about gene multifunctionality in human for the purpose of verifying the relationships between gene multifunctionality and diseases in human. A great body of research has been conducted in the past two decades addressing the similarity between genes using various sources and most commonly using the Gene Ontology (GO) [5, 6]. GO has been extensively used to compute the similarity between genes (*details in section 3*) [19, 20]. In this work, we use the functional annotations of a gene from the Gene Ontology Annotation (GOA) databases to compute the shortest distance (*path length*) between the Molecular Function (*mf*) GO terms annotating the gene. In the GO, molecular function (*mf*) terms are organized as nodes in a tree-like directed acyclic graph (DAG). For example, Figure 1 exhibits 7 nodes representing 7 molecular function terms in GO [5, 6]. Based on the published molecular function annotations of genes from the GO, we examine the relationships between multifunctional genes and diseases and present a computational approach based on path length between *mf* annotations from Gene Ontology as our main metric for estimating the semantics of gene functions and multifunctionality in connection with gene – disease association.

We utilize functional genomics data from OMIM, the Gene Ontology, and GOA Annotation databases and show that multifunctional genes based on *mf* path length approach are significantly more likely to be involved in human diseases compared with other non-multifunctional genes. Thus, we utilized the relationship between the functional terms of a gene to estimate the multifunctionality of the gene. If a gene is multifunctional then it is involved in more than one molecular function in the cell and thus more inclined to be a *disease gene*.

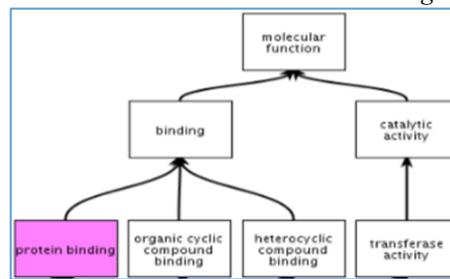

**Figure 1:** a small part of the GO representing 7 nodes from the molecular function aspect.

In other words, we utilized the functional GOA annotations of a gene to predict its multifunctionality and so to determine how likely that gene is associated with diseases. The main *contribution* of this work is in the new technique of determining multifunctional genes using GO *mf* path length which has never been used before as gene multifunctionality metric (*or measure*) to the best of our knowledge. Also we point out here that the relationship between diseases and multifunctional genes is not the main contribution of this paper.

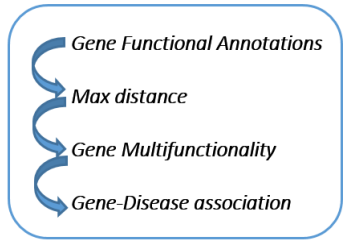

## 2. Background and Related Work

There are several databases and online resources for functions and other information on human genes and the human genome. The NCBI may be regarded as one of the most official resources about human genes in terms of gene name, gene sequence, gene symbols, etc. [17]. For example, when it comes to gene identifier (gene Id) there are at least four gene Id codings as shown in Table 1: NCBI (*enterz*) gene Id, UniprotKB, MIM, …etc [18]. In this work we try to report multiple Id's for each gene.

As reported by several research projects, multifunctional genes tend to have more association with diseases compared with non-multifunctional genes [1–4, 8–10]. Therefor the relationships between diseases and multifunctional genes is signification and proved [1, 10]. A multifunctional gene is a gene that is involved in several functions and activities, including molecular and cellular tasks, inside the cell [1-3, 8]. Typically, studying multifunctional genes can disclose more knowledge about diseases associated with the multifunctional genes. In this paper, we rely on the gene ontology (GO) which is the most popular repository of functional information about human genes [5, 6]. Pritykin, Ghersi, and Singh (2015) presented a comprehensive study of genome-wide multifunctional genes in human [1]. They found that multifunctional genes are significantly more likely to be involved in human disorders [1]. Also, they found that 32% of all multifunctional genes produced by their method are involved in at least one OMIM disorder, whereas the fraction of other annotated genes involved in at least one OMIM disorder is 21% [1, 7].

In [12], Salathe et al. investigated the multifunctionality of yeast genes and proteins for a different goal. They found a positive correlation between how many biological process (*bp*) GO terms a gene is annotated with and its evolutionary conservation in yeast; that is, they found highly significant negative correlation between number of *bp* GO terms and rate of change of yeast genes [12]. Also in [1], they observed that that multifunctional genes tend to be more evolutionarily conserved.

A method for identifying novel moonlighting proteins from current functional annotations in public databases was proposed by Khan et al. (2013) in [3]. They identified potential moonlighting proteins in the Escherichia coli K-12 genome by examining clusters of GO term annotations taken from *UniProt* and constructed three datasets of experimentally confirmed moonlighting proteins [3].

In another study of multifunctional genes [13], Clark and Radivojac (2011) discovered a statistically significant positive correlation between the number of GO biological process leaf terms a gene has and its number of *Pfam* domains and are usually longer [13].

## 3. Methods and Techniques

A great body of research has been conducted in the past two decades addressing the similarity between genes using various sources and most commonly using the *Gene Ontology* (GO) [20]. GO is the main and most comprehensive source of information on gene functions, processes, and cellular localizations [5, 6, 19]. In the past decade or so, it has been used widely and extensively to predict and validate scientific results and findings related to genes [5, 6, 16, 19]. Also, GO has been used extensively to compute the *semantic* similarity between genes using the GO annotation terms of genes [20]. The semantic similarity *measure* is a function that estimates the similarity (or distance) between two genes or two GO terms as a numeric value [19, 20]. Typically, the similarity between two genes will be computed as a function of the similarity between their GO annotation terms.

Basically the similarity $Sim_g(g_1, g_2)$ between two genes $g_1$ and $g_2$ can be a similarity function $Sim_t(.,.)$ between the GO annotation terms of $g_1$ and $g_2$:

$$Sim_g(g_1, g_2) = Sim_t(TS_1, TS_2) \ldots\ldots\ldots\ldots (1)$$

where $Sim_g(g_1, g_2)$ is the similarity between genes $g_1$ and $g_2$; $Sim_t(TS_1, TS_2)$ is the similarity between GO term sets $TS_1$ and $TS_2$; and $TS_1$ and $TS_2$ are the sets of GO terms annotating genes $g_1$ and $g_2$ respectively. That is $TS_i = \{t_1, \ldots, t_n\}$ are the GO terms annotating gene $g_i$.

In this paper we would like to use the path length (*shortest distance*) between GO terms in the molecular function (*mf*) aspect of GO as a metric of gene multifunctionality in connection with gene–diseases relationship. Let $PL(t_1, t_2)$ be the shortest path length between terms $t_1, t_2$ in the GO DAG (*see Figure 2*). For example, in Figure 2, the path length between GO terms (nodes) *A* and *B* is 6, while between *B* and *C* the path length is 3 using edge counting. Let

$$Dist_f(g_k) = \max_{i,j} PL(t_i, t_j) \ldots\ldots\ldots (2)$$

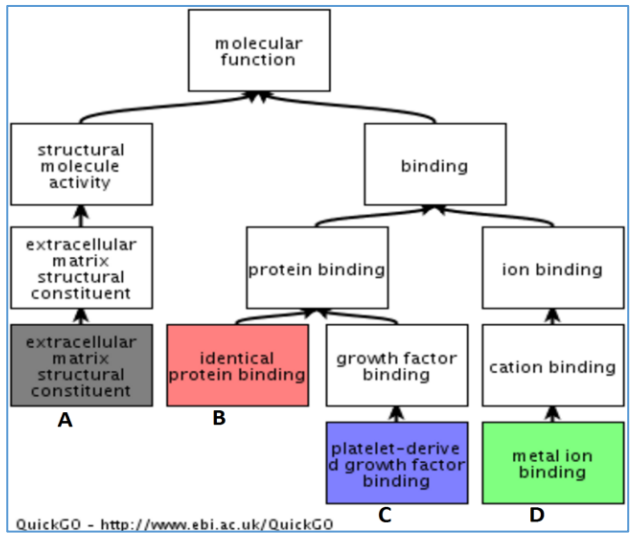

**Figure 2:** Part of the molecular function aspect of the GO showing the four mf annotation terms of gene *COL2A1* in GO {QuickGP – http://www.ebi.ac.uk/QuickGO}

where $t_i$, $t_j$ are two GO annotation terms for gene $g_k$. That is, the *mf* distance $Dist_f(g_k)$ of gene $g_k$ is the maximum pair-wise distance of all *mf* terms of gene $g_k$. For example, in Table 1, we can see that gene *COL2A1* (UniprotKB: P02458; gene Id: 1280; MIM: 120140) has four GO terms (in the molecular function aspect) representing four different functions that this gene is involved in as shown in Figure 2 (nodes *A*, *B*, *C*, and *D*), and these four *mf* GO terms are:

```
GO:0030020 extracellular matrix structural
           constituent conferring tensile strength
GO:0042802 identical protein binding
GO:0046872 metal ion binding
GO:0048407 platelet-derived growth factor binding
```

The distance (shortest path length) between these four *mf* terms, as we can see in Figure 8, ranges from 3 (between *B* and *C*) to 7 (between *A* and *C*). If a gene is involved in two molecular functions that are relatively far apart in the *mf* aspect of the GO then these two functions are highly distinct (different) and more likely the gene is more candidate to be *multifunctional* gene. In this paper we investigate and analyze the maximum distance, equation (2), of all-pair *mf* terms of a gene and its relationship with *multifunctionality* and diseases for human genes.

## 4. Evaluation Results and Discussion

We conducted a number of experimental evaluations on human gene annotations, human diseases, and multifunctionality. We used data from Online Mendelian

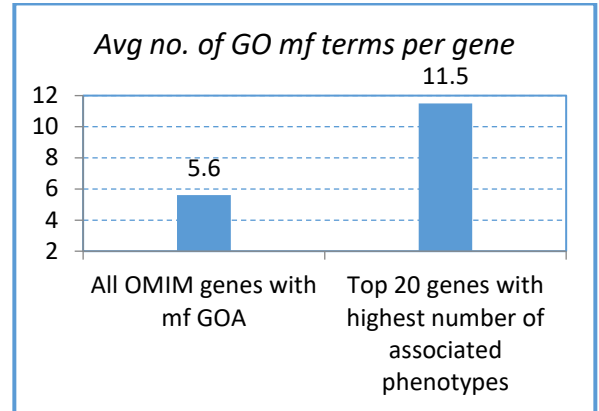

**Figure 3:** On average genes associated with highest number of OMIM diseases have significantly higher number of associated GO *molecular function* (mf) terms (i.e., *11.5 vs. 5.6*) indicating that these genes are likely multifunctional.

Inheritance in Man *OMIM* [7] on gene and disease relationships; also used GOA database for gene functional annotations from the molecular function aspect. We further verified some information and results from *QuickGO* and *NCBI* [17, 21]. In the first experiment, as shown in Figure 3, we examined all human genes available in OMIM [7] that have at least on molecular function *mf* annotation in the gene ontology [5, 6] (roughly >16,000 genes). On average each gene is annotated (associated) with 5.6 molecular functions (*i.e., 5.6 mf terms per gene*). Next, we examined the top 20 genes associated with highest number of diseases according to OMIM and found that on average each one of these genes is annotated with 11.5 molecular functions in the gene ontology as illustrated in Figure 3. This result is significant ($p < 9e\text{-}8$, significance by *hypergeometric test*) which is a clear indication that more molecular functions for a given gene indicate the multifunctionality of the gene and more likelihood of being a disease-gene.

Let a disease-gene be a human gene that is associated with one or more phenotypes in the OMIM database [7], otherwise let us call it a non-disease gene. Next, we proved that disease-genes have significantly higher distance in their molecular function terms in the gene ontology GO (Figure 4) compared with all other genes in the OMIM database [7, 16]. Figure 4 illustrates the significant difference in the GO term (node) distance between mf terms of all genes vs genes associated with highest number of phenotypes in OMIM. It is shown that the maximum distance between all molecular function mf terms in the GO is on average 7.75 for all OMIM genes whereas it is 10.8 for top 20 disease-genes. From these two evaluations, it is very clear that the human genes that are associated with the most number of diseases have significantly higher number of mf terms and higher distances between their mf terms. Next, we investigated the maximum distance between *mf* terms for all genes using the gene ontology annotations GOA. The results are shown in Figure 5. As shown, the largest number of genes (~1400) have distance of 12 between their *mf* terms and the least is distance 2 (~220 genes); Figure 5.

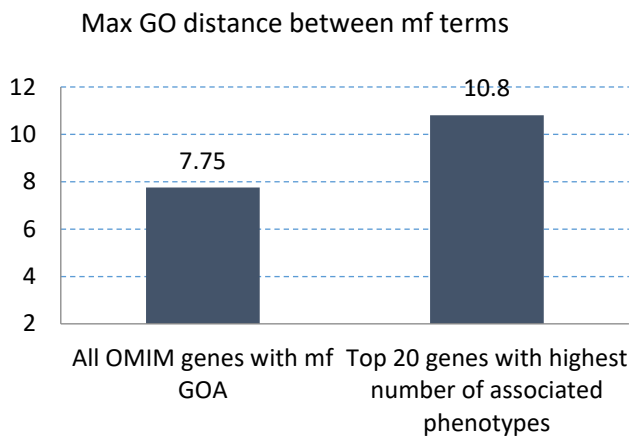

**Figure 4:** On average genes associated with highest number of OMIM phenotypes have higher distance between their mf terms (10.8) compared with all genes (7.75). This makes these genes that are associated with disease more likely to be multifunctional.

To further the study, we examined the total number of *mf* terms and max GO *mf* term distance for the top 20 genes with the highest number of associated diseases. From OMIM, we extracted the top 20 genes having the highest number of phenotypes. These genes are shown in Table 1 (*we included MIM, UniprotKB, and NCBI gene Id for each gene*). We ran our technique and computed number of GO terms and the maximum distance between GO *mf* terms for each one of these 20 genes as shown in Table 1. Evidently, these too heavily disease-associated genes are multifunctional as their *mf* term distance average is 10.8 which is significantly higher than the average of all genes (which is 7.8); see Figure 4 and Table 1. We say multifunctional because each one of these genes is associated on average with 11.5 different molecular functions (*the average is 5.6*).

Moreover, we randomly extracted 20 *non-disease* genes (selected from OMIM randomly). These genes are not associated with any phenotype in OMIM. We analyzed these genes examining their molecular functions (*mf*) in GO and the distance between their *mf* terms and the results are in Table 2 and illustrated in Figure 6. As shown in Table 2, these genes have on average max distance of 6.4 which is not high enough compared with *disease-genes* (average 10.8). In fact, these genes have lower max distance (6.40) than the average of all human genes (which is 7.75 as shown in Figure 4). Moreover, these genes are annotated with an average of 3.6 *mf* terms in the GO which is significantly lower than the average of all genes 5.6 mf terms (*shown in Figure 3*).

Finally, in the Gene Ontology Annotation GOA database, all genes having only one molecular function (GO *mf* term) were extracted and then we randomly selected a set of 20 genes from them shown in Table 3. Thus, these genes are highly likely *non-multi-functional* since each gene is associated with only one *mf* term in the GO. Therefore, we expect these genes to be *non-disease genes* more often than not. This randomly selected set of genes were analyzed and found that only four of them (4 out of 20 or 20%) are associated with phenotypes (*disease genes*) and the remaining (80%) are not (*non-disease genes*). This finding is fairly significant (hypergeometric test, $p < 8e-9$) as the ratio in OMIM database is more than 80% of genes having *mf* annotations are associated with at least one phenotype.

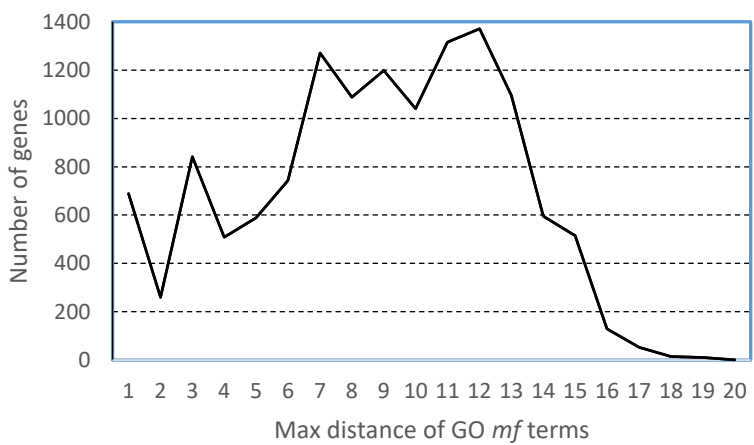

**Figure 5:** Illustration of number of genes for each value of maximum distance between GO mf terms of all OMIM genes.

# 5. Conclusion

This paper presents a method for examining the relationships between diseases from OMIM database and gene ontology annotations of human genes using the molecular function aspect. Specifically, we used the distance between the molecular function *mf* annotation terms as an indicator or metric of gene multifunctionality. Numerous studies have shown that multifunctional genes are more likely associated with diseases compared with non-multifunctional genes [1, 3, 9, 8]. Therefore, we investigated the multifunctionality of a gene using the shortest distance between *mf* GO terms of the gene. The main goal and contribution of this paper is in the way of identifying and determining multifunctional genes using GO *mf* path length and using it as a gene multifunctionality metric which has never been used before to the best of our knowledge. The evaluation results proved that gene multifunctionality indicated by GO term distance is highly correlated with gene–disease association. We showed that, for example, the top 20 genes associated with the highest number of phenotypes in the OMIM database exhibit significantly higher GO distance (*path length*) compared with the average of all human genes in OMIM. We also showed that genes not associated with any disease exhibit significantly lower GO distance compared with all human genes in OMIM; and genes annotated with only one GO *mf* terms have significantly lower disease association compared with all human genes; and more results reported in section 4. All these findings clearly indicate the existence of this interesting correlation. Further study may be needed for a genome-wide correlation investigation with all aspects of GO with all human diseases in OMIM.

Table 1: Top 20 disease genes with highest number of phenotypes in OMIM.

| No. | Gene MIM | UniProtKB | Gen Id | Gene Symbol | Numbr of associated phenotypes | Min mf term distance | Max mf term distance | Nmbr of mf terms |
|---|---|---|---|---|---|---|---|---|
| 1 | 120140 | P02458 | 1280 | COL2A1 | 16 | 3 | 7 | 4 |
| 2 | 134934 | P22607 | 2261 | FGFR3 | 14 | 2 | 11 | 8 |
| 3 | 176943 | P21802 | 2263 | FGFR2 | 14 | 2 | 11 | 10 |
| 4 | 150330 | P02545 | 4000 | LMNA | 12 | 3 | 3 | 2 |
| 5 | 171834 | P42336 | 5290 | PIK3CA | 11 | 1 | 11 | 10 |
| 6 | 190070 | P01116 | 3845 | KRAS | 10 | 2 | 13 | 6 |
| 7 | 300017 | P21333 | 2316 | FLNA | 10 | 1 | 11 | 18 |
| 8 | 120120 | Q02388 | 1294 | COL7A1 | 9 | 1 | 9 | 3 |
| 9 | 134797 | P35555 | 2200 | FBN1 | 9 | 1 | 8 | 8 |
| 10 | 191170 | P04637 | 7157 | TP53 | 9 | 1 | 12 | 31 |
| 11 | 601728 | P60484 | 5728 | PTEN | 9 | 1 | 13 | 17 |
| 12 | 605427 | TRPV4 | 59341 | TRPV4 | 9 | 1 | 13 | 14 |
| 13 | 121014 | P17302 | 2697 | GJA1 | 8 | 1 | 13 | 13 |
| 14 | 139320 | O95467 | 2778 | GNAS | 8 | 1 | 14 | 12 |
| 15 | 141900 | P68871 | 3043 | HBB | 8 | 1 | 9 | 8 |
| 16 | 600163 | Q14524 | 6331 | SCN5A | 8 | 1 | 15 | 17 |
| 17 | 607108 | P26367 | 5080 | PAX6 | 8 | 1 | 14 | 17 |
| 18 | 611731 | P25054 | 324 | APC | 8 | 1 | 10 | 12 |
| 19 | 104311 | P49768 | 5663 | PSEN1 | 7 | 1 | 13 | 15 |
| 20 | 120150 | p02452 | 1277 | COL1A1 | 7 | 1 | 6 | 5 |
| | | | | *Mean:* | **9.7** | **1.35** | **10.8** | **11.5** |

Table 2: Randomly selected 20 *non-disease* genes.

| No. | Gene MIM | UniProtKB | Gene Id | Gene Symbol | Nmbr of mf terms | *Max mf distance* |
|---|---|---|---|---|---|---|
| 1 | 100640 | P00352 | 216 | ALDH1A1 | 5 | *10* |
| 2 | 606380 | Q9BPV8 | 53829 | P2RY13 | 1 | *0* |
| 3 | 607097 | Q9NSA0 | 55867 | SLC22A11 | 4 | *8* |
| 4 | 603555 | P63272 | 6827 | SUPT4H1 | 4 | *8* |
| 5 | 600295 | P16860 | 4879 | NPPB | 3 | *2* |
| 6 | 604637 | O95158 | 11247 | NXPH4 | 2 | *3* |
| 7 | 154360 | O43451 | 8972 | MGAM | 6 | *9* |
| 8 | 610808 | A0A087X179 | 10272-3859 | TBC1D3E | 1 | *0* |
| 9 | 605664 | O95235 | 10112 | KIF20A | 6 | *13* |
| 10 | 015504 | O15504 | 11097 | NUPL2 | 4 | *8* |
| 11 | 164060 | P55209 | 4673 | NAP1L1 | 2 | *5* |
| 12 | 602223 | Q13541 | 1978 | EIF4EBP1 | 4 | *8* |
| 13 | 115442 | P19784 | 1459 | CSNK2A2 | 4 | *11* |
| 14 | 600187 | P63241 | 1984 | EIF5A | 7 | *8* |
| 15 | 607020 | Q8WWW0 | 83593 | RASSF5 | 3 | *8* |
| 16 | 602500 | Q14789 | 2804 | GOLGB1 | 4 | *7* |
| 17 | 611811 | Q96JL9 | 84449 | ZNF333 | 2 | *6* |
| 18 | 600141 | P61604 | 3336 | HSPE1 | 6 | *7* |
| 19 | 607698 | Q96JN0 | 84458 | LCOR | 1 | *0* |
| 20 | 185860 | Q7L0J3 | 9900 | SV2A | 3 | *7* |
| | | | | *Average* | **3.6** | **6.40** |

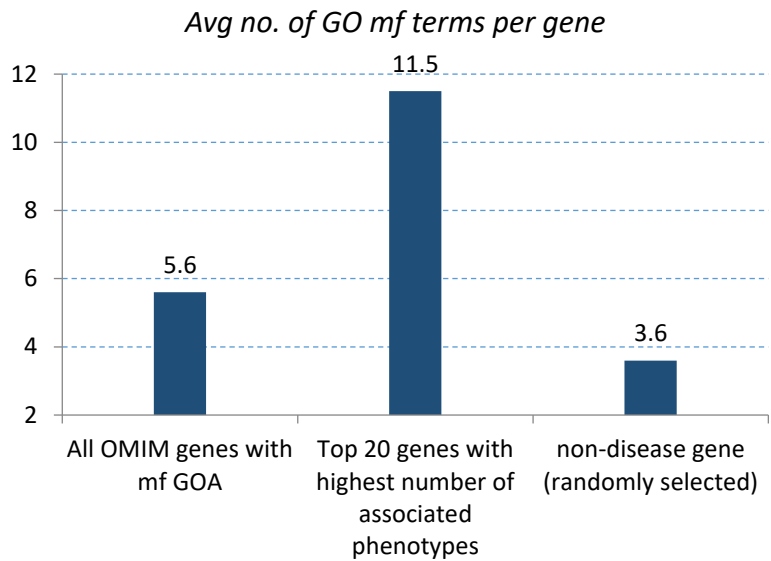

**Figure 6:** Illustration of different types of human genes compared by their associated number of molecular functions mf from the gene ontology GO.

**Table 3:** A list of 20 genes randomly selected from all human genes having only one *mf* GO term (~3900 genes).

| Seq. | MIM | Gene Id | Gene symbol | Pheno-types | Notes |
|---|---|---|---|---|---|
| 1 | 300618 | 9767 | JADE3 | 0 | *non-disease-gene* |
| 2 | 605006 | 23401 | FRAT2 | 0 | *non-disease-gene* |
| 3 | 603835 | 4705 | NDUFA10 | 1 | *disease-gene* |
| 4 | 613993 | 58498 | MYL7 | 0 | *non-disease-gene* |
| 5 | 610337 | 55665 | URGCP | 0 | *non-disease-gene* |
| 6 | 603199 | 10207 | PATJ | 0 | *non-disease-gene* |
| 7 | 612435 | 28232 | SLCO3A1 | 0 | *non-disease-gene* |
| 8 | 608680 | 51002 | TPRKB | 0 | *non-disease-gene* |
| 9 | 611000 | 26122 | EPC2 | 0 | *non-disease-gene* |
| 10 | 120220 | 1291 | COL6A1 | 1 | *disease-gene* |
| 11 | 146790 | 2212 | FCGR2A | 1 | *disease-gene* |
| 12 | 605783 | 10286 | BCAS2 | 0 | *non-disease-gene* |
| 13 | 602051 | 5301 | PIN1P1 | 0 | *non-disease-gene* |
| 14 | 300794 | 441519 | CT45A3 | 0 | *non-disease-gene* |
| 15 | 607051 | 147323 | STARD6 | 0 | *non-disease-gene* |
| 16 | 602502 | 2800 | GOLGA1 | 0 | *non-disease-gene* |
| 17 | 142987 | 3231 | HOXD1 | 0 | *non-disease-gene* |
| 18 | 602908 | 10741 | RBBP9 | 0 | *non-disease-gene* |
| 19 | 607139 | 2175 | FANCA | 0 | *disease-gene* |
| 20 | 609933 | 130120 | REG3G | 0 | *non-disease-gene* |